\documentclass[12pt,draft]{article}
 
\usepackage{amsmath}
\usepackage{amsthm}
\usepackage{amssymb}
\usepackage{epsf}

\begin{document}

\thispagestyle{empty}
\rightline{HUB-EP-99/03}
\rightline{MIT-CTP-2823}
\rightline{hep-th/9901041}
\vspace{2truecm}
\centerline{\bf \Large New N=1 Superconformal Field Theories}
\vspace{.5truecm}
\centerline{\bf \Large and
their Supergravity Description}

\vspace{1.5truecm}
\centerline{\bf 
Andreas Karch\footnote{karch@ctp.mit.edu}}
\vspace{.4truecm}
{\em \centerline{Center for Theoretical Physics, MIT, 
Cambridge, MA 02139, USA}}             
\vspace{.5truecm}
\centerline{\bf Dieter L\"ust\footnote{luest@physik.hu-berlin.de}\ and \
Andr\'e Miemiec\footnote{miemiec@physik.hu-berlin.de}}

\vspace{.4truecm}
{\em 
\centerline{Humboldt-Universit\"at, Institut f\"ur Physik,
D-10115 Berlin, Germany}}

\vspace{1.0truecm}
\begin{abstract}
In this note we construct a new class of superconformal field
theories as mass deformed $N=4$ super Yang-Mills theories.
We will argue that these theories 
correspond to the fixed points which were
recently found \cite{warner} studying the   deformations
of the dual IIB string theory on $AdS_5\times S^5$.
\end{abstract}

\bigskip \bigskip
\newpage



\section{Introduction}

The investigation of superconformal field theories has already a long history.
One appealing feature of superconformal models is that they often
exhibit a strong-weak coupling duality symmetry (S-duality).
In a remarkable recent development it became clear that in a large class
of superconformal field theories a new type of duality symmetry arises,
namely  they can be equivalently described by supergravity
in anti-de-Sitter (AdS) spaces \cite{maldacena}. 
In particular, there is a correspondence
between four-dimensional superconformal field theories and
supergravity on $AdS_5\times M^5$, where $M^5$ is a certain five-dimensional
Einstein space. In the simplest case, $M^5$ is given by $S^5$, and the
corresponding superconformal field theory is just $N=4$ super Yang-Mills
with $SU(n)$ gauge symmetry. This is nothing else as
the superconformal theory which
lives on the world volume of $n$ parallel D3 branes.
Another well studied example for $M^5$ is the coset space $T^{1,1}$
which leads to a $N=1$ superconformal gauge theory, which is the celebrated
superconformal theory of D3 branes probing a conifold singularity 
\cite{klewit}.

The prescription \cite{gubklebpol,wit,hensken,gubser} 
of the holographic map allows for
several non-trivial checks of the conjectured AdS/CFT correspondence.
For example, the central charge of the conformal field theory is 
inversely proportional to the volume of $M^5$.
This check works very nicely
for the correspondence
between the coset space $T^{1,1}$ and the superconformal field theory
from D3 branes at the conifold singularity.
Recently, 
deformations of
the usual IIB string theory on $AdS_5 \times S^5$ has been
studied \cite{zaff,dist,warner} by analyzing critical points of 
$N=8$ gauged supergravity. In 
\cite{warner} a new fixed point was found, leaving $N=2$ unbroken
in the bulk (corresponding to $N=1$ on the brane).
The information
about this fixed point that has been extracted from the 
supergravity description
are the global symmetries and the ratio of the central charges of the 
undeformed
UV theory and the interacting IR theory: $c_{IR}/c_{UV}=27/32$.
The aim of this letter is to show that this new fixed point corresponds
to a particular mass deformation of the $N=4$ super-Yang-Mills theory.
We will show
that our new fixed point field theories obtained by
mass deforming the $N=4$ theory indeed 
reproduce the global symmetry as well as the ratio $c_{IR}/c_{UV}$
from the supergravity side.

In section 2 we will first briefly review 
the method 
\cite{leighstrassler} of deforming 
a supersymmetric field theory with a marginal
operator to obtain a new class of superconformal models.
Then we recall
the existence of a family
of $N=1$ superconformal theories as mass deformed $N=2$ theories.
Deforming the finite $N=2$ $SU(n)$ theory with $2n$ flavors by a
mass for the adjoint chiral multiplet leaves an $N=1$ theory with
a quartic superpotential. By the method of \cite{leighstrassler} it
can be established that this quartic superpotential is a marginal
deformation of the IR physics. The superconformal
theory along the fixed line parametrized by the marginal operator
is precisely the superconformal theory of $n$ D3 branes
probing a conifold.


Then we will argue that the same arguments will basically establish
that deforming the $N=4$ SUSY gauge theory by a mass term to one of the 
adjoint chiral multiplets will lead to a one parameter 
family of $N=1$ superconformal theories. They can be expressed as $N=1$
theories with two massless adjoints $A$ and $B$ deformed
by a quartic superpotential
$W\sim (AB)^2$.
This has to be contrasted with the mass deformation of the $N=4$ theories
by a mass for a full hypermultiplet studied e.g. in \cite{donagiwitten}.
While latter one leaves an $N=2$ theory, our deformation leaves
only $N=1$ unbroken.

In section 3 we will turn to the dual supergravity description
provided by the supersymmetric fixed point found in \cite{warner} from the
deformation of the $AdS_5\times S^5$ supergravity. While the
field theory considerations presented up to this point are actually valid for
an arbitrary gauge group, only the $SU(n)$ theories will be
realized on D3 branes probes in IIB\footnote{Allowing
for orientifolds, the $SO$ and $Sp$ examples are also accessible.}.

\section{The new conformal theories}

Let us first briefly recall the construction of 
\cite{leighstrassler} to establish the existence 
of a family of  new $N=1$ superconformal theories by mass deforming 
a given superconformal theory. The mass deformation causes a flow 
from the original theory in the
UV to the deformed theory in the IR.
Specifically, start with a theory at a fixed point with a marginal
operator provided by the superpotential
\begin{equation}
W=g X \phi \phi',
\end{equation}
and add the following mass term to the superpotential
\begin{equation}
W_{\rm mass}=m X^2.
\end{equation}
Via its equation of motion the heavy field can be integrated out, and one
obtains the new, non-renormalizable superpotential
\begin{equation}
W_{\rm new}=-\frac{g^2}{2m}(\phi\phi')^2.
\end{equation}
As shown in \cite{leighstrassler} this is again a marginal operator.

\subsection{The flow from finite $N=2$ theories to superconformal 
$N=1$ theories}

In this section be briefly recall how this method 
establishes the existence and S-duality of a family
of $N=1$ superconformal theories as mass deformed finite $N=2$ theories.
In the simplest case the $N=2$ theory is $SU(n)$ gauge theory with $n$
fundamental hypermultiplets, whose $\beta$-function is zero.
In $N=1$ language the superpotential has the form
\begin{equation}
W=g Q \tilde Q X,
\end{equation}
where $X$ is an adjoint scalar multiplet and the $2n$ 
fundamental fields $Q$ and $\tilde Q$
originate from the hypermultiplets. 
Now giving mass to $X$ via $W_{\rm mass}=mX^2$ breaks the supersymmetry to
$N=1$ with, after integrating out $X$, the marginal operator
\begin{equation}
W_{\rm new}=h(Q\tilde Q)^2,\quad h=\frac{g^2}{2m}.
\end{equation}

As discussed in \cite{leighstrassler} there is a second way to flow to the
curve parametrized by this operator.
They consider supersymmetric quantum chromesodynamics (SQCMD),
that is $SU(n)$ gauge theory with $2n$ flavors, a singlet meson
field $N$ and:
$$
W= \lambda N Q \tilde{Q} + \frac{m_0}{2} N^2.
$$
Again there is the
 marginal operator\footnote{Keeping track of the index structure
the second term in the superpotential
should really read $N^s_r N^r_s - \frac{1}{N} N^r_r N^s_s$
for $n$ colors. Similarly the $(Q\tilde Q)^2$ operator will read
$(Q^r_{\alpha} \tilde{Q}^{\alpha}_s)
(Q^s_{\beta} \tilde{Q}^{\beta}_r) - \frac{1}{N}
(Q^r_{\alpha} \tilde{Q}^{\alpha}_r)
(Q^s_{\beta} \tilde{Q}^{\beta}_s)$.
Here and in the rest of the paper we will use the compact
and sloppy notation $(Q\tilde Q)^2$.} $(Q\tilde Q)^2$,
since
$ \beta_{gauge} \propto (n-n \gamma_Q) \propto \beta_{\lambda}$
and hence vanishing of the $\beta$ functions only imposes
one constraint on the two couplings.
Integrating out $N$ from the superpotential, we generate the marginal operator
with a coupling $\lambda/2m_0$.

One interesting point is that the S-duality of the finite 
$N=2$ theory, $g\leftrightarrow \frac{1}{g}$, translates directly into
a $N=1$ S-duality, $h\leftrightarrow \frac{1}{h}$, on the fixed line
parametrized by $h$.
The special point where this operator is turned off corresponds
to $N=1$ with no superpotential. In SQCMD this can be achieved at 
$m_0=\infty$, 
in $N=2$
by $g=0$. 
At this point the global symmetry is enhanced. 
In SQCMD there is another special point with this enhanced symmetry,
$m_0=0$. Here we have Seiberg's dual $SU(n)$ with $2n$ flavors and 
$W=Nq\tilde q$.
In the $N=2$ language this corresponds to the free
magnetic theory at $g=\infty$.
Therefore the $N=2$ S-duality means from
  the $N=1$ point of
view  that the theory is selfdual under Seiberg duality.

This type of  $N=1$ gauge theory 
with gauge group $SU(n)\times SU(n)$, bifundamental chiral matter
fields and with quartic superpotential precisely appears as the superconformal
theory living on $n$ D3 branes probing a conifold,
respectively as the dual supergravity on $AdS_5\times T^{1,1}$
($T^{1,1}=(SU(2)\times SU(2))/U(1))$. In a T-dual brane
picture \cite{uranga,mukhi} \`a la Hanany-Witten \cite{hw}, 
the mass deformations corresponds to the
rotation of one of the two NS branes by a certain angle. On the other hand, 
the flow from $N=2$ to $N=1$ corresponds in the supergravity
context to deforming the $N=2$ orbifold space $S^5/Z_2$ by a blow up
to the coset space $T^{1,1}$.

\subsection{The flow from $N=4$ theories to superconformal $N=1$ theories}

By the same reasoning as in the finite $N=2$ case we can also
study the $N=4$ theory. This is really just a special case of a
finite $N=2$ theory. The matter content is just 
one adjoint hypermultiplet. So the
analysis from above applies to this case as well. This has
however some interesting implications.
So let us spell out this ``result'' that is implicit in the analysis
of \cite{leighstrassler}.

From the $N=1$ point of view, the $N=4$ theory provides us
three adjoint chiral fields $A$, $B$ and $X$. 
The superpotential is just the cubic expression $W=gf^{abc}A_aB_bX_c$.
Now we add the mass term for the chiral field $X$. As a result we get that
any $N=1$ theory with two adjoint matter fields $A$ and $B$ allows
for a marginal deformation by adding the quartic superpotential
\begin{equation}
W=\frac{g^2}{2m}f^{abc}f^{dec}{A_aB_bA_dB_e}.\label{marop}
\end{equation}
Not all values of this marginal coupling are distinct. There
exists an S-duality inherited from the $N=4$ theory, mapping strong
coupling to weak coupling.\\ This is difficult to see from
the field theory point of view, but it is a direct consequence of type
IIB S-duality and the AdS/CFT correspondence
once the dual supergravity description is established.
As in the case of the finite $N=2$ theory this S-duality seems to imply
a selfduality of the 2 adjoint theories under Seiberg duality.
\noindent
Let us add a few comments about this model:

\vskip0.2cm
\noindent (i)
The self-duality  under
Seiberg duality is not in apparent conflict with the models with
two adjoints discussed by \cite{brodiestrassler} where always an ADE-type
of superpotential is present.

\vskip0.2cm
\noindent
(ii) In contrast to the previous case we cannot reach the marginal operator 
eq.(\ref{marop}) from an SQCMD description with four singlet meson 
fields\footnote{We are grateful 
to M. Strassler for correcting us on this point.}.
Four massive singlets would yield
\begin{equation}
W = ({\rm tr} A B)^2
\end{equation}
which can only be identified with (\ref{marop}) if the
product of two $f^{abc}$ symbols can be written as a product of
$\delta^{ab}$ symbols. This is however only possible for $SU(2)$.

\vskip0.2cm
\noindent (iii)
This deformation is not the same as the deformation
of the $N=4$ theory by a mass term for a full hypermultiplet as e.g.
studied in 
\cite{donagiwitten}. Latter one leaves an $N=2$ SUSY unbroken and the
mass deformation is given by one complex parameter. In our case
it is just one real mass parameter.

\vskip0.2cm
\noindent (iv)
As in the
previous case of deforming $N=2$
models there will be again
a description in terms of deformed  brane configurations, now in terms of a 
brane box. In contrast
to the realization of the mass deformed $N=4$ theory on the interval
studied by \cite{witten} where one gives a complex mass to a hypermultiplet
leaving $N=2$ unbroken, the brane box naturally give the possibility
to incorporate a real mass for a chiral multiplet
(breaking down to $N=1$) by a very similar
mechanism. A more detailed description of this duality will be given
in \cite{workinpro}.

\section{The dual supergravity description}

Following the ideas of \cite{maldacena} one would expect
these conformal field theories to have a dual supergravity description.
Since the field theory arises as a mass deformation of $N=4$ SYM,
the dual supergravity description should be a deformation of
the usual IIB string theory on $AdS_5 \times S^5$. In \cite{zaff,dist,
warner} such deformations where studied by analyzing critical points of 
$N=8$ gauged supergravity. In 
\cite{warner} a new fixed point was found, leaving $N=2$ unbroken
in the bulk (corresponding to $N=1$ on the brane). We will argue that
this deformation indeed corresponds to the dual of the superconformal
field theories we were studying in this paper.

As a first piece of evidence for our identification of the conformal
$N=1$ theory obtained by mass deforming the $N=4$ theory with
the SUGRA solution of 
\cite{warner} let us compare the global symmetries.
According to \cite{warner} the subgroup of $SO(5)$ unbroken
by the solution is $SU(2) \times
U(1)$. The $SU(2)$ in the field theory rotates the 2 adjoints into
each other. The $ABAB$ super potential is invariant. 
The $U(1)$ is the $U(1)_R$ symmetry of
the $N=1$ theory under which $A$ and $B$ both have charge $1/2$.

A more quantitative test is to compare the ratio of the central charges $c$
of the undeformed UV and the deformed IR theory. 
The UV central charge
will be given just by the free field contributions.
The central charge of the IR conformal theory can be calculated from
the anomaly of the R-charge, since they sit in the same
supermultiplet. This calculation can be found
in great detail e.g. in \cite{gubser}.

On the supergravity side the ratio can be calculated by comparing the volume
of the undeformed SUGRA solution
with that of the deformed. The prediction is 27/32.
Let us show that this value is reproduced by our proposed dual
field theory.

As in \cite{gubser} we calculate the central charge $c$ and the
axial charge $a$ computing \cite{anselmi}
the correlators among the energy momentum
tensor $T$ and the R-current $R$:
\begin{equation}
\partial_\mu <TTR>~\sim ~ a-c 
\end{equation}
and, with the same proportionality factor,
\begin{equation}
\frac{9}{16}  \partial_\mu <RRR>~\sim ~ 5a - 3c.
\end{equation}

First consider the UV theory. The UV theory is the unbroken $N=4$ theory. 
It has $c=1/4\cdot (N_c^2-1)$.
To see this use the above relations.
We have $(N_c^2-1)$ gauginos with $r=1$ and
      $3\cdot (N_c^2-1)$ matter fermions with $r=-1/3$ (the 
      superpotential has to have $r=2$,
so the scalars have $r=2/3$ and the fermions $r=-1/3$).

$\partial_\mu <TTR>$ is given by the sum of all $r$-charges,
\begin{equation}
  \partial_\mu <TTR>=  (N_c^2-1)\cdot
                       \left[\vbox{\vspace{2ex}\hbox{1}\vspace{0ex}}\right] 
                      +3\cdot (N_c^2-1) \cdot 
                       \left[-\frac{1}{3}\right]  = 0
\end{equation}
hence $a-c=0$ or $a=c$.
Moreover
\begin{eqnarray}
     \frac{9}{16} \partial_\mu <RRR> &=&\frac{9}{16}
        \left\{ 
                  (N_c^2-1) \cdot 
                  \left[\vbox{\vspace{2ex}\hbox{1}\vspace{0ex}}\right]^3 
                + 3\cdot (N_c^2-1)\cdot 
                  \left[-\frac{1}{3}\right]^3
        \right\}\nonumber\\
                                     &=&\frac{1}{2}\cdot (N_c^2-1)
\end{eqnarray}
hence $5a -3c =\frac{1}{2}\cdot (N_c^2-1)$ and with $a=c$ we get 
\begin{equation}
    c_{UV} = \frac{1}{4}\cdot (N_c^2-1) .
\end{equation}
     
In the IR we see the mass deformed $N=1$ theory,
so $W=ABX + X^2$ produces $W=(AB)^2$,
and we are left with
$(N_c^2-1)$ gauginos with $r$-charge $r=1$ plus
    $2\cdot (N_c^2-1)$ matter fermions $A,B$ with $r$-charge $r=-1/2$.
Therefore
\begin{equation}
    \partial_\mu <TTR>~=~  (N_c^2-1)\cdot
                           \left[\vbox{\vspace{2ex}\hbox{1}}\right]
                          +2\cdot (N_c^2-1) \cdot \left[-\frac{1}{2}\right] 
                           = 0,
\end{equation}
 and hence still $a=c$.
In addition we have
\begin{eqnarray}
    \frac{9}{16} \partial_\mu <RRR> &=& \frac{9}{16}
                \left\{
                     (N_c^2-1)\cdot 
                     \left[\vbox{\vspace{2ex}\hbox{1}}\right]^3 
                   + 2\cdot (N_c^2-1)\cdot 
                     \left[-\frac{1}{2}\right]^3
                 \right\} \nonumber\\
                                    &=&\frac{27}{64}\cdot (N_c^2-1)~,
\end{eqnarray}
and hence $5a-3c=2c= \frac{27}{64}\cdot (N_c^2-1)$ or
\begin{equation}
      c_{IR} =\frac{27}{128}\cdot (N_c^2-1) .
\end{equation}

Comparing with above we find $c_{IR}/c_{UV} =27/32$
as predicted by supergravity!
It would be interesting to compare also the chiral spectrum
of the superconformal field theory with the spectrum of the scalar
Laplacian of the deformed $S^5$ manifold. 

Note that the numerical value $27/32$ is precisely the same as the one
obtained in the related setup of the conifold as viewed as a mass deformation
of the $Z_2$ orbifold 
\cite{gubser,witten}. This lead \cite{warner} to the speculation that
these two theories are indeed related. Here we see that they are quite
distinct. The reason for the matching of the numerical values is just
due to the mechanism by which we deform: a finite theory with a cubic
superpotential (the only choice in a finite theory) is deformed
by a mass term, giving rise to quartic superpotential while killing
1/3 of the fields. The superpotential uniquely fixes the $r$-charge which
in turn determines the central charge. 

Having identified the deformation of \cite{warner} leading to
an $N=1$ superconformal field theory in the dual language, one
might hope that we can understand the deformations
leading to the $N=0$ theories in a similar spirit.

\vskip0.5cm
\noindent{\bf Acknowledgements:}

\noindent 
 Work partially supported by the E.C. project
ERBFMRXCT960090 and the Deutsche Forschungs Gemeinschaft.
We like to thank M. Aganagic for  useful discussions. We are grateful
to M. Strassler for correcting an error in an earlier version.

\end{document}